\documentclass[aps,showpacs,twocolumn]{revtex4}
\usepackage{amsmath}
\usepackage{amsfonts}
\usepackage{amssymb}
\usepackage{graphicx}

\usepackage{indentfirst}

\begin{document}
\title{Strong field double ionization~: The phase space perspective}

\author{F. Mauger$^1$, C. Chandre$^2$, T. Uzer$^3$}
\affiliation{$^1$ Ecole Centrale de Marseille, Technop\^ole de Ch\^ateau-Gombert, 38 rue Fr\'ed\'eric Joliot Curie
F-13451 Marseille Cedex 20, France
\\$^2$ Centre de Physique Th\'eorique, CNRS -- Aix-Marseille Universit\'es, Campus de Luminy, case 907, F-13288 Marseille cedex 09, France \\ $^3$ School of Physics, Georgia Institute of Technology, Atlanta, GA 30332-0430, USA}
\date{\today}

\begin{abstract}
We identify the phase-space structures that regulate atomic double ionization in strong ultrashort laser pulses. The emerging dynamical picture complements the recollision scenario by clarifying the distinct roles played by the recolliding and core electrons, and leads to verifiable predictions on the characteristic features of the ``knee'', a hallmark of the nonsequential process.
\end{abstract}
\pacs{32.80.Rm, 05.45.Ac}
\maketitle

\begin{figure}
        \includegraphics[width=80mm]{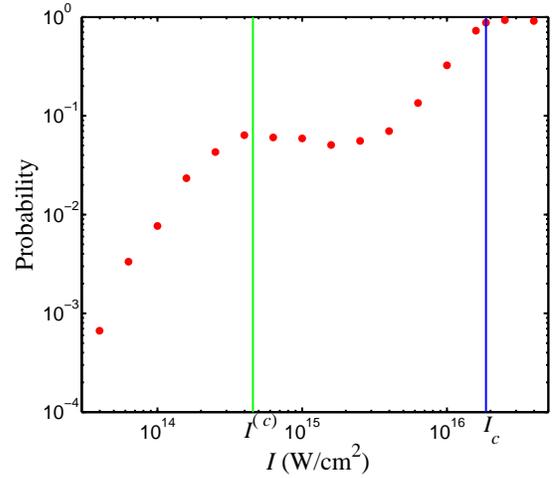}
        \caption {\label{fig:1}   
        Double ionization probability for Hamiltonian~(\ref{Hamiltonian}) as a function of the intensity of the field $I$ for $\omega=0.0584$ a.u. 
        The vertical lines indicate (in green) the laser intensity $I^{(c)}= 4.57 \times 10^{14}\ \mbox{W} \cdot \mbox{cm}^{-2}$ where our dynamical analysis predicts the maximum of nonsequential double ionization, and (in blue) the intensity $I_c= 1.86 \times \ 10^{16}\ \mbox{W} \cdot \mbox{cm}^{-2}$ where the double ionization is expected to be complete (see Fig.~\ref{fig:10}).}   
\end{figure}

One of the most striking surprises of recent years in intense laser-matter interactions has come from multiple ionization by intense short laser pulses:  Correlated (nonsequential) double ionization rates were found to be 
several orders of magnitude higher than the uncorrelated 
sequential mechanism allows. This discrepancy has made the characteristic ``knee'' shape in the 
double ionization yield versus intensity plot into one of the 
most dramatic manifestations of electron-electron correlation in nature. 
The precise mechanism that makes correlation so effective is far 
from settled. Different scenarios 
have been proposed to explain the mechanism behind ionization~\cite{fitti92,cork93,scha93,walk94,beck96,kopo00,lein00,sach01,fu01,panf01,barn03,colg04,ho05_1,ho05_2,ruiz05,horn07,prau07,feis08} and have been confronted with experiments~\cite{brya06,webe00_2}, the recollision 
scenario~\cite{cork93,scha93}, in which the ionized electron is hurled back at the ion core by the laser, being in best accord with experiments. 
In Fig.~\ref{fig:1}, a typical double ionization probability as a 
function of the intensity of the laser field is plotted.
Similar knees have been observed in 
experimental data~\cite{fitti92,kodo93,walk94,laro98,webe00_2,corn00,guo01,dewi01,ruda04} and successfully 
reproduced by quantal computations on atoms and molecules~\cite{beck96,wats97,lapp98,panf03}.
In a recent series of articles ~\cite{fu01,sach01,panf02,panf03,ho05_1,ho05_2,liu07} characteristic 
features of double ionization were reproduced using classical trajectories
and this success was ascribed to the paramount role of correlation~\cite{ho05_1}. Indeed, entirely classical interactions turn out to be adequate to generate the strong two-electron correlation needed for 
double ionization. 

In this Letter, we complement the well-known recollision scenario by identifying the organizing principles which explain the statistical properties of the classical trajectories such as ionization probabilities. In addition to the dynamical picture of the ionized electron provided by the recollision scenario, we connect the dynamics of the core electron and the energy flow leading to double ionization to relevant phase space structures (periodic orbits or invariant tori). The resulting picture leads to two verifiable predictions for key points which make up the knee  
in Fig.~\ref{fig:1}: namely the laser intensity where nonsequential double ionization is maximal and the intensity where the double ionization is complete.

We work with the classical Hamiltonian model of the helium atom with soft Coulomb 
potentials~\cite{su91,java88}. The Hamiltonian is given 
by~\cite{panf01}:
\begin{eqnarray}
    && {\mathcal H}(x,y,p_{x}, p_{y},t) =  \frac{p_{x}^{2}}{2} + \frac{ 
p_{y}^{2}}{2}+(x+y)E(t) \nonumber \\
    &&  \quad                       +\frac{1}{\sqrt{(x-y)^{2}+1}}
                             -\frac{2}{\sqrt{x^{2}+1}} 
-\frac{2}{\sqrt{y^{2}+1}},
                             \label{Hamiltonian}
\end{eqnarray}
where $x$, $y$ and $p_{x}$, $p_{y}$ are the positions and 
(canonically conjugate) momenta of each electron respectively.
The energy is initially fixed at the ground state ${\mathcal E}_g=-2.24$~a.u.~\cite{haan94}.
The laser field is modeled by a sinusoidal pulse with an envelope, i.e.\ 
$E(t)= E_{0} \ f(t) \ \sin \omega t$ where $E_{0}$ is the maximum amplitude and $\omega$ the laser frequency chosen as 
$\omega=0.0584$~a.u.~which corresponds to a wavelength of 780~nm. The 
pulse envelope $f(t)$ is chosen as a trapezoidal function with 2-4-2 
laser pulse shape (the ramp-up lasts two cycles, the plateau four, and 
the ramp-down two)~\cite{ho05_1,ho05_2,panf03,panf02}. 
Typical ionizing trajectories of Hamiltonian~(\ref{Hamiltonian}) show two 
qualitatively different routes to double ionization~: 
nonsequential double ionization (NSDI), where the two electrons leave 
the core (inner) region at about the same time, and sequential double 
ionization (SDI), where one electron leaves the inner region long time 
after the other one has ionized.

\begin{figure}
        \includegraphics[width=80mm]{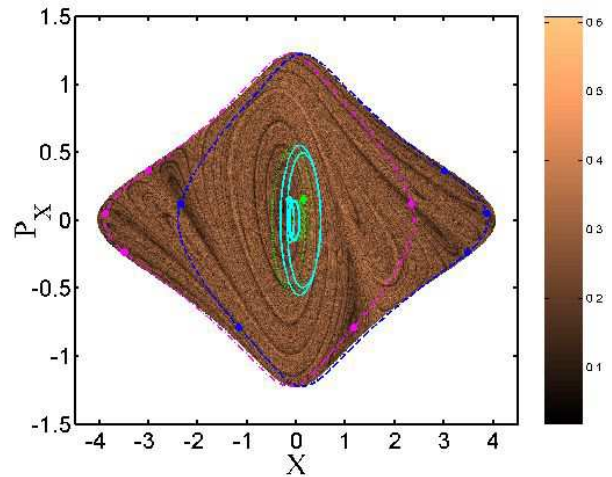}
        \caption{\label{fig:5}
        FTL maps of Hamiltonian~(\ref{Hamiltonian}) without the field at time $t=43$ a.u.~in the plane $(x,p_{x})$ with $y=0$. The projection of some important periodic orbits (as continuous curves) and their respective Poincar\'e sections (as dots)~:
$O_{x,1}$ dashed dotted line (pink online),
$O_{x,2}$ dashed line (blue online),
$O_{y,1}$ dotted line (green online),
$O_{y,2}$ full line (cyan online).
        }
\end{figure}

We first analyze the dynamics 
of Hamiltonian~(\ref{Hamiltonian}) without the field ($E_0=0$) using 
linear stability properties such as obtained by the finite-time Lyapunov (FTL) 
exponents ~\cite{chaosbook}. FTL maps quantify the degree of chaos in phase space and highlight invariant structures. With each initial condition of a particular ensemble (here 
taken on the plane $(x,p_x)$ with $y=0$, the other coordinate $p_y$ 
being determined by the energy condition ${\mathcal H}={\mathcal E}_g$), we associate the coefficient 
$\log|\lambda(t)|/t$ where $\lambda(t)$ is the largest 
eigenvalue of the Jacobian obtained at time $t$ from the tangent 
flow~\cite{chaosbook}. 
A typical FTL map is depicted in Fig.~\ref{fig:5} 
for Hamiltonian~(\ref{Hamiltonian}) without the field. It clearly 
displays strong and global chaos by showing fine details of the 
stretching and folding of trajectories~\cite{chaosbook}. In particular, there are no
regular elliptic islands of stability contrary to what is common with Hamiltonian 
systems on a bounded energy manifold.
By examining typical trajectories, we notice that the motion of the two 
electrons tracks, at different times, one of four hyperbolic periodic orbits, denoted~$O_{x,1}$, 
$O_{x,2}$, $O_{y,1}$ and~$O_{y,2}$ and displayed in Fig.~\ref{fig:5}. Their period is 29~a.u., i.e., much shorter than the 
duration of the laser pulse (of order 800 a.u.). These four orbits are symmetrical [reflecting the symmetries of the Hamiltonian~(\ref{Hamiltonian})]~: The 
two periodic orbits~$O_{x,1}$ and $O_{x,2}$ [outer projections in the plane~$(x,p_x)$]
have on the plane $(y,p_y)$, the same projections as the periodic 
orbits $O_{y,1}$ and $O_{y,2}$ [inner projections in the plane $(x,p_x)$]. Consequently, a typical two-electron trajectory is composed of one electron close to the nucleus (the ``inner'' electron) and another 
further away (the ``outer'' electron), with quick exchanges of the roles of each electron. We will see below that this distinction is crucial: Since the contribution of the field-electron 
interaction to Hamiltonian~(\ref{Hamiltonian}) is proportional to the 
position, the action of the field is larger for the outer electron, 
while the inner electron is mainly driven by the interaction with the 
nucleus. 

\begin{figure}
        \includegraphics[width=80mm]{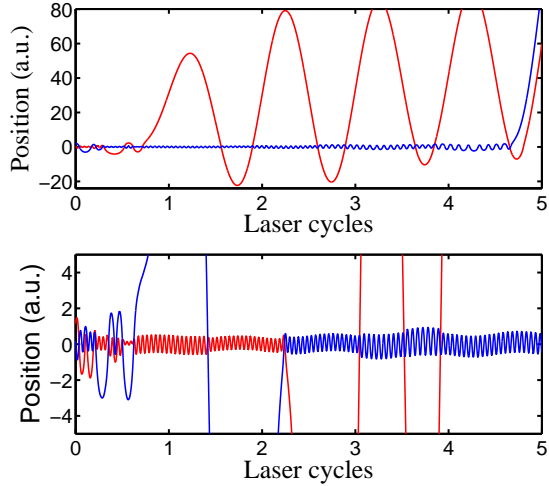}
        \caption{\label{fig:7}
        Two typical trajectories of Hamiltonian~(\ref{Hamiltonian}) for $I=10^{15}\ \mbox{W} \cdot \mbox{cm}^{-2}$ for initial conditions in the ground state energy of the helium atom. The two positions ($x$ in red and $y$ in blue) are plotted versus time (expressed in laser cycles). Note the vastly different vertical scales of the two panels. The recollision mechanism is seen in both panels~: In the upper one, the recollision (at the end of the panel) brings in enough energy to ionize the inner electron. In the lower panel, the recollision energy insufficient to ionize the inner electron -- the electrons exchange roles instead.
        }
\end{figure}

\paragraph*{Single ionization--}

By switching on the field, the outer electron is picked up and swept away from the 
nucleus. Consequently, its effective Hamiltonian is~:
\begin{equation} \label{seq:H1}
{\mathcal H}_1=\frac{p_x^2}{2} + E_0 x f(t) \sin\omega t.
\end{equation}
We notice that Hamiltonian ${\mathcal H}_1$ is integrable. Its solutions are approximately composed of linear escape from the nucleus (at time $t_0$) modulated by the action of the field ~\cite{cork93,beck94,band05} (see the red trajectory in Fig.~\ref{fig:7}). 

For the inner electron, the effective Hamiltonian contains the interaction with the nucleus and with the laser field~:
\begin{equation} \label{Ham:H2}
{\mathcal H}_2=\frac{p_y^2}{2}-\frac{2}{\sqrt{y^2+1}}+yE_0\sin\omega t.
\end{equation}

In the absence of the field ($E_0=0$), ${\mathcal H}_2$ is also integrable and the inner electron is confined on a periodic orbit. Since it stays close to the nucleus, its approximate period is $2\pi/\sqrt{2}$ obtained from the harmonic approximation, as observed in Fig.~\ref{fig:7}.

\paragraph*{Sequential double ionization (SDI)--}

\begin{figure}
       \includegraphics[width=80mm]{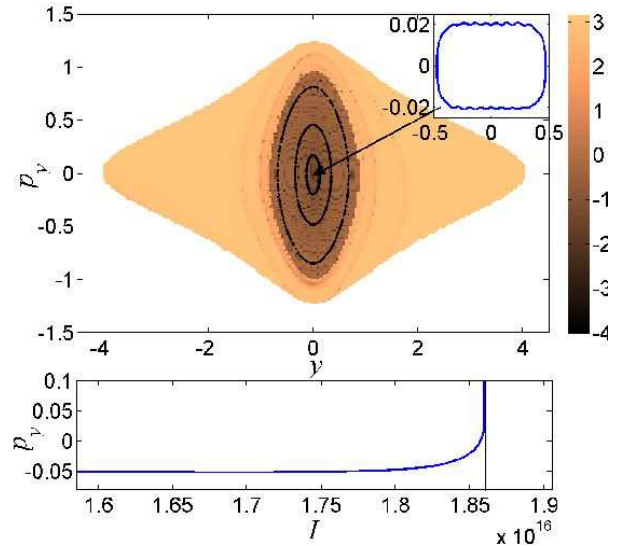}
       \caption{\label{fig:10}
       Upper panel: Contour plot of the electron location $y(t)$ at time $t=215.2$~a.u. (2 laser cycles), of
Hamiltonian~(\ref{Ham:H2}) for $I=5\times 10^{15}\ \mbox{W} \cdot \mbox{cm}^{-2}$. Poincar\'e
sections (stroboscopic plot) of selected trajectories in the elliptic central region are
also depicted. The color code is on a logarithmic scale. The inset shows a projection of this periodic orbit at $I=1.7 \times 10^{16} \ \mbox{W} \cdot \mbox{cm}^{-2}$ in the $(y,p_y)$-plane.
Lower panel~: Momentum of the central periodic orbit (on the Poincar\'e section) of 
Hamiltonian~(\ref{Ham:H2}) as a function of the laser intensity. The vertical line on the lower panel indicates the intensity $I_c=1.86\times 10^{16} \ \mbox{W} \cdot \mbox{cm}^{-2}$ such that for $I\geq I_c $, complete unhindered SDI is expected. 
       }
\end{figure}

Once an electron has been ionized (usually during the ramp-up of the field), the other electron is left with the nucleus and the field. Its effective Hamiltonian is ${\mathcal H}_2$. A contour plot of the electron excursions after two laser cycles and a Poincar\'e section of ${\mathcal H}_2$ are depicted in Fig.~\ref{fig:10} for $I=5\times 10^{15}\ \mbox{W} \cdot \mbox{cm}^{-2}$. They clearly show two distinct regions~: 
The first one is the core region which is composed of a collection of invariant tori which are slight deformations of the ones obtained in the integrable case ${\mathcal H}_2$ without the field. This elliptic region is organized around a main elliptic periodic orbit which has the same period as the field $2\pi/\omega\approx 107.6$~a.u. In this region, the electrons are fairly insensitive to the field, and do not ionize. The second region is the one outside the core where trajectories ionize quickly. It corresponds to sequential double ionization.  In between these two regions, any weak interaction (with the outer electron for instance) may move the inner electron confined on the outermost regular tori (but still inside the brown elliptical region) to the outer region where it ionizes quickly. 

If the laser intensity $I$ is too small, then the phase space is filled with invariant tori and no sequential double ionization can occur because the motion is regular. The sequential double ionization probability depends then on the size of the regular region around the elliptic periodic orbit, and hence on $I$.
We have computed the location and the stability of this periodic orbit for $\omega=0.0584$~a.u. using a Newton-Raphson algorithm~\cite{chaosbook}. When it exists, this periodic orbit stays elliptic in the whole range of intensities we have considered. On the stroboscopic plot (with frequency $\omega$) the periodic orbit is located at $y=0$. In Fig.~\ref{fig:10}, the momentum $p_y$ of the periodic orbit on the stroboscopic plot is represented as a function of $I$. We notice that for a large set of intensities in the range $[10^{14}, 10^{16}] \mbox{W} \cdot \mbox{cm}^{-2}$, this periodic orbit is located close to $p_y= 0$. For intensities larger than $I_c=1.86 \times 10^{16}\ \mbox{W} \cdot \mbox{cm}^{-2}$, the periodic orbit does not exist, and no major islands of regularity remain. Therefore, it is expected that the sequential double ionization probability is equal to one in this range
of intensities, as observed on the probability curve on Fig.~\ref{fig:1}.

\paragraph*{Nonsequential double ionization (NSDI)--}

As noted before, when the field is turned on, its action is concentrated on only one electron, the outer one, as a first step.  The field drives the outer electron away from the nucleus, leaving the inner electron nearly unaffected by the field because its position remains small. 
From the recollision process~\cite{cork93,scha93}, the outer electron might come back close to the nucleus during the pulse plateau, if the field amplitude is not too large. In this case, it transfers a part of its energy to the inner electron through the electron-electron interaction term. 
From then on, two outcomes are possible~: If the energy brought in by the outer electron is sufficient for the other electron to escape from the regular region (as in Fig.~\ref{fig:7}, upper panel), then it might ionize together with the outer electron. The maximum energy ${\mathcal E}_x$ of the outer electron when it returns to the inner region (after having left the inner region with a small momentum $p_0$ close to zero) is obtained from Hamiltonian~(\ref{seq:H1}) and is
${\mathcal E}_x= \kappa U_p$, where $U_p=E_0/(4\omega^2)$ is the ponderomotive energy and $\kappa= 3.17...$ is the maximum recollision kinetic energy in units of $U_p$~\cite{cork93,beck94,band05}. We complement the recollision scenario (which focuses on the outer electron) by providing the phase space picture of the inner electron~: In order to ionize the core electron, the energy brought back by the outer electron has to be of order of the energy difference between the core ($y=0$) and the boundary of the stable region ($y=y_m$) of ${\mathcal H}_2$ (see Fig.~\ref{fig:10}) which is equal to
\begin{equation}
\label{eq:DE}
\Delta {\mathcal E}_y=2-\frac{2}{\sqrt{y_m^2+1}}.
\end{equation}
A good approximation to $y_m=y_m(E_0)$ is given by the value where the potential is locally maximum,~i.e.\ $E_0= 2y_m/(y_m^2+1)^{3/2}$.
The equal-sharing relation which links the classical picture of the outer electron $x$ with the one of the inner electron $y$,
\begin{equation}
\label{eq:DEEx}
\Delta {\mathcal E}_y=\frac{{\mathcal E}_x}{2}= \frac{\kappa}{2\omega^2}\frac{y_m^2}{(y_m^2+1)^3},
\end{equation} 
defines (through an implicit equation) the expected value of the field $E_0^{(c)}$ for maximal NSDI, because it describes the case when each outer electron brings back enough energy to ionize the inner electron, while remaining ionized itself. However, fulfilling this energy requirement does not guarantee NSDI: The outcome depends on the number and efficiency of recollisions. For $\omega=0.0584$, the predicted value of the amplitude $E_0^{(c)}$ as given by Eq.~(\ref{eq:DEEx}) corresponds to an intensity of $I^{(c)}=4.57\times 10^{14}\ \mbox{W} \cdot \mbox{cm}^{-2}$ which agrees very well with the simulations shown in Fig.~\ref{fig:1}.
In a wide range of frequencies, an accurate expansion of $E_0^{(c)}$ is obtained from Eqs.~(\ref{eq:DE})-(\ref{eq:DEEx}) and given by
\begin{equation}
\label{eq:E0c}
E_0^{(c)}= \frac{4\omega}{\sqrt{\kappa}}-\left(\frac{2\omega}{\sqrt{\kappa}} \right)^{3/2}+O\left(\frac{4\omega^2}{\kappa}\right),
\end{equation}
for sufficiently small $\omega$. To leading order the corresponding intensity varies as $\omega^2$. For $\omega=0.0584$, the approximate intensity given by Eq.~(\ref{eq:E0c}) is $4.60\times 10^{14}\ \mbox{W} \cdot \mbox{cm}^{-2}$ which is in excellent agreement with $I^{(c)}$.
When the field $E_0$ is too small, then the outer electron cannot gain enough energy to ionize the inner electron. When the field $E_0$ is too large, then the outer electron does not recollide since it leaves the interaction region nearly linearly. These two limits explain the bell shape of the resulting nonsequential double ionization probability, which, when put together with the monotonic rise of the SDI probability at higher intensities, adds up to the knee in question.
 
\acknowledgments
CC acknowledges financial support from the PICS program of the CNRS. This work is partially funded by NSF.



\end{document}